\documentclass[aps,prx,twocolumn,superscriptaddress,nofootinbib]{revtex4-2}
\usepackage{graphicx,amssymb,amsmath,amsfonts,empheq,latexsym,braket,bm,bbm,dsfont,upgreek}
\usepackage[normalem]{ulem}
\usepackage[dvipsnames]{xcolor}
\usepackage{comment}
\usepackage{tabularx}
\newcolumntype{C}{>{\centering\arraybackslash}X}
\usepackage{color}
\usepackage[dvipsnames]{xcolor}
\usepackage{makecell}
\usepackage{multirow}
\usepackage{url}
\usepackage{soul}
\usepackage{algorithm}
\usepackage{algpseudocode}
\usepackage{hyperref}
\hypersetup{
colorlinks = true,
linkcolor = [rgb]{0,0,0}, 
citecolor = [rgb]{0.13,0.55,0.13},
urlcolor = [rgb]{0.25, 0.41, 0.88}}

\usepackage{tikz}
\usepackage{tikz-cd}
\usetikzlibrary{arrows}
\usetikzlibrary{intersections}
\usetikzlibrary{shapes.geometric}
\usetikzlibrary{decorations.pathmorphing, patterns,shapes}
\usetikzlibrary{decorations.markings}

\tikzset{
	mid arrow/.style={postaction={decorate,decoration={
				markings,
				mark=at position .575 with {\arrow[#1]{stealth}}
	}}},
	near arrow/.style={postaction={decorate,decoration={
				markings,
				mark=at position .275 with {\arrow[#1]{stealth}}
	}}},
	far arrow/.style={postaction={decorate,decoration={
				markings,
				mark=at position .800 with {\arrow[#1]{stealth}}
	}}},
}

\renewcommand{\leq}{\leqslant}
\renewcommand{\geq}{\geqslant}

\newcommand{\bbI}{\mathbb{I}}

\newcommand{\bbZ}{\mathbb{Z}}

\newcommand{\calC}{\mathcal{C}}

\newcommand{\calL}{\mathcal{L}}
\newcommand{\calM}{\mathcal{M}}

\newcommand{\calO}{\mathcal{O}}
\newcommand{\calP}{\mathcal{P}}

\newcommand{\calT}{\mathcal{T}}

\newcommand{\eqnref}[1]{Eq.~\eqref{#1}}
\newcommand{\figref}[1]{Fig.~\ref{#1}}

\newcommand{\tabref}[1]{Table~\ref{#1}}
\newcommand{\secref}[1]{Sec.~\ref{#1}}

\begin{document}

\title{Simulating the non-unitary Yang-Lee conformal field theory on the fuzzy sphere}

\author{Ruihua Fan}
\affiliation{Department of Physics, University of California, Berkeley, CA 94720, USA}
\author{Junkai Dong}
\affiliation{Department of Physics, Harvard University, Cambridge, MA 02138, USA}
\affiliation{Kavli Institute for Theoretical Physics, University of California, Santa Barbara, California 93106, USA}
\author{Ashvin Vishwanath}
\affiliation{Department of Physics, Harvard University, Cambridge, MA 02138, USA}

\begin{abstract}

The fuzzy sphere method has enjoyed great success in the study of (2+1)-dimensional unitary conformal field theories (CFTs) by regularizing them as quantum Hall transitions on the sphere.
Here, we extend this approach to the Yang-Lee CFT—the simplest non-unitary CFT.  
We use an Ising quantum-Hall ferromagnet Hamiltonian with a transverse field and an imaginary longitudinal field, the latter breaks the Hermiticity of the Hamiltonian and thus the unitarity of the associated quantum field theory.
Non-unitary conformal field theories—particularly the Yang-Lee CFT—pose significant  challenges to conventional fuzzy sphere approaches. To overcome these obstacles, here we utilize a different method for determining critical points that requires no a priori knowledge of CFT scaling dimensions. Our method instead leverages the state-operator correspondence while utilizing two complementary criteria: the conformality of the energy spectrum and its consistency with conformal perturbation theory.
We also discuss a new finite-size scaling on the fuzzy sphere that allows us to extract conformal data more reliably, and compare it with the conventional analysis using the (1+1)-dimensional Yang-Lee problem as an example.
Our results show broad agreement with previous Monte-Carlo and conformal bootstrap results. We also uncover one previously unknown primary operator and several operator product expansion coefficients.

\end{abstract}

\maketitle

\section{Introduction}

Conformal field theories (CFTs) are local quantum field theories with translation, rotation/Lorentz, and importantly, scale invariance~\cite{Ginsparg:1988ui,Christe:1993ij,DiFrancesco:1997nk,Rychkov:2016iqz,Simmons-Duffin:2016gjk}.
Due to their widespread occurrence, understanding CFTs is a central pursuit in both theoretical and experimental physics~\cite{Aharony:1999ti,Cardy:1996xt,Sachdev:2011fcc,dhar2006theoretical,Calabrese:2005in}.
CFTs, by definition,  lack intrinsic energy scales, rendering conventional perturbation inapplicable. Furthermore, most CFTs in three and higher spacetime dimensions are strongly interacting. Therefore, non-perturbative approaches become necessary.

The most successful method so far is the conformal bootstrap, which is exceptionally powerful in two dimensions and also efficient for unitary CFTs in higher dimensions~\cite{Belavin:1984vu,El-Showk:2012cjh,Poland:2018epd}.
An alternative method is to numerically simulate the desired CFTs via critical points of microscopic models.
Typically, such a model is defined on a lattice to ensure a numerically manageable finite-dimensional Hilbert space. Therefore the model cannot have exact continuous spatial symmetries, which introduces various irrelevant perturbations at the critical point exacerbating the finite size effect. 
Moreover, CFT data is usually extracting by calculating correlation functions.
The difficulty of expressing generic primary operators in the CFT in terms of microscopic lattice operators makes the task of exhausting all the low-lying primaries a formidable one.

Recently, a new method has been proposed to study (2+1)-dimensional CFTs, the fuzzy sphere regularization, which partially overcomes these challenges~\cite{Madore:1991bw,Hasebe:2010vp,Zhu:2022gjc,Zhou:2023qfi,Hofmann:2023llr,O(3)FuzzySphere:2023,zhou2024newseries3dcfts,Voinea:2024ryq,3dpotts}.
It uses the lowest Landau level (LLL) on the two-dimensional sphere to create microscopic models with an exact $SO(3)$ rotation symmetry and a finite-dimensional Hilbert space simultaneously~\cite{Haldane:1983xm,Greiter2011}.
Specifically, the model comprises multilayer LLLs at an integer filling so that the charge fluctuation remains gapped while the layer pseudospin undergoes quantum phase transitions~\cite{Sondhi1993,Zaletel:2018HalfFilledLL}.
Thanks to the state-operator correspondence, the entire CFT operator content can be determined by the energy spectrum at criticality, obtained via exact diagonalization, allowing access to many more (including previously undiscovered) primary operators.
Importantly, this method does not rely on the unitarity of the CFT or the Hermiticity of the Hamiltonian, making it applicable to nonunitary CFTs as well, for which the bootstrap method is not efficient.

In this work, we apply the fuzzy sphere method to examine one of the simplest non-unitary CFTs, the (2+1)-dimensional Yang-Lee CFT/singularity~\cite{YangLee1,YangLee2,Cardy:2023lha}. 
Our microscopic model is a natural extension of the one previously used for the Ising CFT~\cite{Zhu:2022gjc}. However, it suffers from pronounced finite-size effects, likely due to the combined influence of the small scaling dimension of the lowest primary operator and the non-Hermitian nature of the Hamiltonian.
Here, we present a systematic approach to control these effects and perform reliable finite-size scaling. This method takes advantage of the structural features of the fuzzy sphere and is different from conventional finite-size scaling.
We use the (1+1)-dimensional Yang-Lee problem as an example to draw a comparison. 

Using this approach, we obtain a well-resolved conformal spectrum and can reliably extract the CFT data.
There have been various estimates for the scaling dimension of the lowest primary operator~\cite{Fisher1978,deAlcantaraBonfim:1981sy,2005LatticeAnimal,Butera:2012tq}.
Recent bootstrap studies using truncation methods have revealed information about two additional primaries and operator product expansion (OPE) coefficients~\cite{Gliozzi:2013ysa,Gliozzi:2014jsa,Hikami:2017hwv}.
Our results not only confirm these findings but also identify a new spin-2 primary operator and extract several previously unknown OPE coefficients.

The remainder of the manuscript is organized as follows. 
\secref{sec:model} introduces the microscopic Hamiltonian, its phase diagram, and briefly reviews the basics of Yang-Lee CFTs. 
The reader primarily interested in the extracted conformal data may wish to proceed directly to \secref{sec:result}.
\secref{sec:FSS} discusses the finite-size scaling method tailored to the fuzzy sphere.
In \secref{sec:comparison}, we apply the conventional scaling approach to both the (1+1)D Yang-Lee lattice model and the (2+1)D fuzzy sphere system, illustrating the subtleties and motivating the necessity for the modified method.
We conclude with a discussion and possible future directions in \secref{sec:discussions}.

\section{The model and phase diagram}
\label{sec:model}

Consider the bilayer quantum Hall problem on a two-dimensional sphere. For simplicity, we work with the unit $|e| = \hbar = c = 1$.
The magnetic field on the sphere, after the LLL projection, sets the short-distance length scale and is normalized to unity. 
The radius of the sphere $R$ determines the number of fluxes $N_\phi = 2 R^2$ and should also correspond to the radius perceived by the infrared quantum field theory. 
We focus on the unit filling, at which the number of electrons $N$ and the radius $R$ are related by $R = \sqrt{(N-1)/2}$.
Use $c_{a}(\bm{r})$ as the LLL-projected fermion annihilation operator on layer $a= \uparrow,\downarrow$, and introduce three fermion bilinears 
\begin{equation}
	n_\alpha = \begin{pmatrix} c_\uparrow^\dag & c_\downarrow^\dag \end{pmatrix} \sigma_\alpha \begin{pmatrix} c_\uparrow \\ c_\downarrow \end{pmatrix}\,,\quad \alpha = x,y,z
\end{equation} 
where $\sigma_\alpha$ is the Pauli matrix.
We refer to $n_\alpha$ as pseudospin variables, whose low-energy fluctuation will constitute the degrees of freedom of the target CFT.

The Hamiltonian for realizing the Yang-Lee CFT comprises three terms, a transverse field, an imaginary longitudinal field, and a ferromagnetic interaction
\begin{equation}
\label{eq:YL fuzzy}
\begin{aligned}
	H = & - \int d^2 \bm{r} n_x(\bm{r}) + i\lambda n_z(\bm{r}) \\
	& \quad + \int d^2 \bm{r} d^2 \bm{r}' V(\bm{r}-\bm{r}') n_\uparrow (\bm{r}) n_\downarrow(\bm{r}') 
\end{aligned}
\end{equation}
where $V(\bm{r}-\bm{r}')$ involves a finite number of pseudo-potentials. 
The imaginary longitudinal field renders the Hamiltonian non-Hermitian and the associated quantum field theory non-unitary.
Below, we fix the transverse field as our energy unit, use the longitudinal field and pseudopotentials to drive the transition.

\begin{figure}
\centering
\begin{tikzpicture}
\draw[thick, ->, >=stealth] (-0.1,0) -- (3.7,0) node[right]{$V$};
\draw[thick, ->, >=stealth] (0,-0.1) -- (0,3.) node[left]{$\lambda$};
\filldraw[red!60] (2.8,0) circle (0.05) node[black,below]{\scriptsize 2+1D Ising};
\filldraw[blue!80] (0,2) circle (0.05) node[black, above right]{\scriptsize 0+1D};
\node at (0.8,1.85) {\scriptsize Yang-Lee};
\node at (-0.5,2) {\scriptsize $\lambda=1$};
\draw (0,2) .. controls (0.2,1.5) and (1.5,0.2) .. (2.8,0);
\node at (0.9,0.55) {\scriptsize $PT$-};
\node at (0.9,0.3) {\scriptsize symmetric};
\node at (2.2,1.2) {\scriptsize $PT$-broken};
\end{tikzpicture}
\caption{A schematic phase diagram of the Yang-Lee Hamiltonian \eqnref{eq:YL fuzzy}. The horizontal axis stands for the strength of the ferromagnetic interaction not any specific pseudopotential. In the $V=0$ limit, different LLL orbitals are decoupled and the system undergoes a 0+1D Yang-Lee phase transition at $\lambda = 1$.}
\label{fig:phase diagram}
\end{figure}
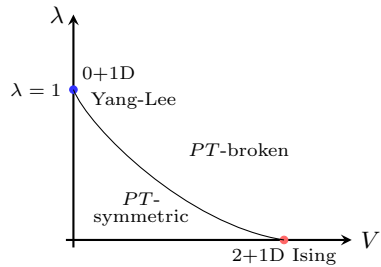

The longitudinal field breaks the Ising $\bbZ_2$ symmetry. A non-unitary $\calP\calT$ transformation, which combines the pseudospin flip and complex conjugation, commutes with the Hamiltonian:
\begin{equation}
	\calP\calT: n_z \rightarrow - n_z\,,\, i \rightarrow -i\,.
\end{equation}
If the $\calP\calT$ symmetry is unbroken, the ground state and all low-energy states will have real eigenenergies despite the Hamiltonian's non-Hermitian nature~\cite{Bender:2023cem}. The Yang-Lee CFT is associated with the spontaneous breaking of this $\calP\calT$ symmetry. 

The Hamiltonian also maintains the $SO(3)$ rotation symmetry and the spatial reflection symmetry. In particular, the reflection across the $xz$ plane of the pseudospin variables $n_\alpha(\bm{r})$ can be generated by a combination of layer exchange and charge conjugation.
Specifically, consider
\begin{equation}
	\calP\calC: \begin{pmatrix} c_{\uparrow} \\ c_{\downarrow} \end{pmatrix} \rightarrow i\sigma_y  \begin{pmatrix} c_{\uparrow}^\dag \\ c_{\downarrow}^\dag \end{pmatrix}\,,\, i \rightarrow i\,.
\end{equation}
Let $(\theta,\varphi)$ denote the polar and azimuthal angles of the spherical coordinates. 
One can use the property of the LLL wavefunctions to verify
\begin{equation}
	\calP\calC: n_\alpha(\theta,\varphi) \rightarrow n_\alpha(\theta,-\varphi)\,.
\end{equation}
One arrives at the same conclusion that $\calP\calC$ generates the improper $O(3)$ transformation by examining how $\calP\calC$ acts on an $SO(3)$ vector in the orbital space~\cite{Zhu:2022gjc}.

\figref{fig:phase diagram} shows the ground state phase diagram schematically.
At $\lambda = 0$, the problem is reduced to the Ising phase transition. The paramagnetic and ferromagnetic phases smoothly connect to the $\calP\calT$-symmetric and $\calP\calT$-broken phases, respectively. The Yang-Lee CFT sits on the entire phase boundary.
In the 3D Euclidean space, we can capture the Yang-Lee CFT by a non-unitary Landau-Ginzburg Lagrangian
\begin{equation}
    \calL = (\partial \phi)^2 + i (g - g_c) \phi + i g_{\phi^3} \phi^3\,,
    \label{eq:landau ginzberg}
\end{equation} 
where $\phi$ is a $\calP\calT$-odd real scalar field~\cite{Fisher1978}.
As a consequence of the cubic interaction, $\phi^2$ becomes a descendant of $\phi$, which leaves $\phi$ as the only relevant primary operator. 
This is consistent with having a critical line in the phase diagram.
We expect the following fusion rule
\begin{equation}
	[\phi] \times [\phi] = [\bbI] + [\phi] + [\phi^3] + [T_{\mu\nu}] + \cdots\,,
\end{equation}
where $\bbI$ is the conformal vacuum, $\phi^3$ the leading irrelevant primary scalar and $T_{\mu\nu}$ the stress-energy tensor.\footnote{At the fixed point $\phi$ and $\phi^3$ should be regarded only as abstract symbols of the corresponding primaries, and it is not meaningful to think of $\phi^3$ as the cube of $\phi$.}

In terms of the energy spectrum, the ground state energy is real in the $\calP\calT$-symmetric phase and at the phase boundary, and it becomes imaginary in the other phase.
The onset of imaginary ground state energy provides one diagnosis of the phase transition, and it is especially useful for studying the one-dimensional lattice version of the Hamiltonian~\cite{vonGehlen:1991zlm}. 
For the (2+1)D problem, we will adopt a more efficient method.

\section{Finite-size scaling on the fuzzy sphere}
\label{sec:FSS}

All numerical simulation methods of quantum critical points face two technical challenges: locating the critical point and mitigating finite-size effects. 
The conventional wisdom is to use the finite-size scaling of dimensionless quantities and increase the system size with fixed parameters.
The fuzzy sphere has some unique features and subtleties that call for alternative methods.
In this section, we explain our method of determining the critical point and how to properly perform the finite-size scaling analysis.

\subsection{Determine the critical point on the fuzzy sphere}

On the fuzzy sphere, having state-operator correspondence and a handful of tuning parameters (the pseudopotentials) provides us with a more efficient method of determining the critical point.
We directly apply the gradient descent to find the Hamiltonian parameter that produces an energy spectrum with approximate integer spacing.
Namely, we take the spectrum having conformality as the {\em definition} of the finite-size system being critical. 
As we shall see, the conventional way of identifying the critical point via finite-size scaling of dimensionless quantities will generally fail to agree with this method for subtleties that we discuss in the next subsection.

For the current problem, we choose the pseudopotentials and the longitudinal field $\lambda_z$ as the tuning parameters at a given system size.
Without assuming any specific knowledge of the operator scaling dimensions in the target CFT, we choose the cost function for the gradient descent to be
\begin{equation}
\begin{aligned}
	f(\{V\},\lambda_z) = \min_{\alpha>0} \sum_{O,n} & (\alpha\,\delta E_{O,n}^{\text{Num}} - \delta E_{O,n}^{\text{CFT}})^2 \\
    & + (\alpha\,E_{T_{\mu\nu}}^{\text{Num}} - E_{T_{\mu\nu}}^{\text{CFT}})^2\,,
    \label{eq:conformality cost function}
\end{aligned}
\end{equation}
where $\alpha$ is a non-universal rescaling factor, the upperscript Num means the quantity is obtained by numerics. 
Here, $E_{O, n}$ is the energy of the $n$-th descendant in the conformal family $[O]$ and $\delta E_{O,n} = E_{O, n} - E_{O, 0}$ the energy difference. 
In practice, we find it sufficient to choose the first six excited states that correspond to $\phi,\partial_\mu\phi, \Box \phi, \partial_\mu\partial_\nu\phi$, $\partial_\mu \Box \phi$ and the stress-energy tensor $T_{\mu\nu}$. 
Since $\phi$ should violate the unitarity bound $\Delta_\phi < 1/2$, we know that the five $[\phi]$ states are the lowest-energy states in the corresponding angular momentum sectors, and $T_{\mu\nu}$ is the second state in the angular momentum $\ell = 2$ sector. 
This method has also been used in Ref.~\cite{zhou2024newseries3dcfts} to search for new CFTs.

In practice, we find it necessary to fine tune four pseudopotentials to obtain a reasonable conformal spectrum.
The gradient descent in such a five-dimensional parameter space might yield many unwanted local minima. We can then use conformal perturbation theory to examine the validity of the output parameter~\cite{Cardy:1989da, Lao:2023zis, Lauchli:2025fii}.
Namely, the microscopic Hamiltonian \eqref{eq:YL fuzzy} near the true critical point can be captured by the following effective form
\begin{equation}
	\label{eq:perturbation Hamiltonian}
	H = \frac{v}{R} H_{\text{CFT}} + i \int \big( g_\phi \phi + g_{\phi^3} \phi^3 + \ldots)\, d^2 x\,,
\end{equation}
where $H_{\text{CFT}}$ is the CFT Hamiltonian, $v$ is a non-universal number, $g_{\phi}$ and $g_{\phi^3}$ are coefficients of the relevant and leading irrelevant perturbation, ellipsis represents other irrelevant terms. We pick our convention of the sign of the OPE coefficient $C_{\phi\phi\phi}$ so that the $\calP\calT$-symmetric phase corresponds to $g_\phi > 0$.
If the output parameter corresponds to the genuine Yang-Lee CFT, we can then apply conformal perturbation theory (CPT) to \eqnref{eq:perturbation Hamiltonian} to match {\em all} the energy spectra near this parameter~\cite{Lao:2023zis, Lauchli:2025fii}. 
For simplicity, we consider the correction only from $\phi$ to the first five states in the conformal family of $[O]$ and the stress tensor $T_{\mu\nu}$.
We choose the ratio of the OPE coefficients $r = C_{\phi \phi \phi}/C_{TT\phi}$ as the variational parameters to minimize the following cost function by gradient descent
\begin{equation}
	f_{\{V\},\lambda_z}(r) = \min_{\alpha>0, g_\phi} \sum_i ( E_i^{\text{CPT}} + \delta E_i^{\text{CPT}} - \alpha E_i^{\text{Num}} )^2
\end{equation}
at \emph{every} point near the putative critical point.
On the right-hand side, $\Delta_i$ is the CFT scaling dimension extracted through the spectrum at the output parameter, and $\delta E_i^{\text{CPT}} \propto g_\phi$ is analytically calculated via conformal perturbation.
The output parameter corresponding to the putative critical point should at least approximately produce a vanishing relevant perturbation $g_\phi=0$.
Once the output parameter passes the validity check, it should indeed give rise to a CFT Hamiltonian with small irrelevant perturbations.
One can then obtain a reasonable estimate of the scaling dimensions and OPE coefficients using the optimized parameter at this given system size.

As a side remark, in principle, we can apply the conformal perturbation calculation to extract the scaling dimension $\Delta_\phi$ as well by treat it as another variational parameter. One generally will find that the result does not exactly agree with that obtained from the conformal spectrum. Since doing the first order perturbation and ignoring higher irrelevant operators introduces errors, one should treat this as a benchmark instead of a more accurate approach to read out the conformal data.

\subsection{Increasing system sizes on the fuzzy sphere}

In principle, increasing the system size is beneficial for further reducing the effect of residual irrelevant perturbations and giving more accurate results.
However, an important subtlety on the fuzzy sphere renders the naive finite-size scaling invalid. 
Specifically, given the same interaction $V(\bm{r}-\bm{r}')$ in the real space, the pseudopotential expansion coefficients $V_J$ in the angular momentum space at different radii $R$ differ asymptotically by $\calO(R^{-2})$, and vice versa~\cite{Zhu:2022gjc}.
This ambiguity in the definition of pseudopotentials, which disappears only in the thermodynamic limit, can cause the optimal parameters to have a small drift with the system size~\cite{Voinea:2024ryq,Lauchli:2025fii}.

We can quantify this drift for the Yang-Lee problem by relating the microscopic Hamiltonian \eqref{eq:YL fuzzy} to the effective Hamiltonian \eqref{eq:perturbation Hamiltonian} more concretely.
The lack of an on-site unitary symmetry implies that any microscopic operator can overlap with both $\phi$ and $\phi^3$.
Therefore, $g_{\phi}$ and $g_{\phi^3}$ in \eqref{eq:perturbation Hamiltonian} should be linearly related to longitudinal fields and pseudopotentials near the critical point.
Let $\lambda_{z,c}$ and $V_c$ denote the critical coupling in the thermodynamic limit. In a system at radius $R$, we have the following ansatz
\begin{equation}
\begin{aligned}
    g_\phi =& a_\phi (\lambda_{z} - \lambda_{z,c}) + b_\phi (V - V_c) + \frac{c_\phi}{R^2} + \ldots \\
    g_{\phi^3} =& a_{\phi^3} (\lambda_z - \lambda_{z,c}) + b_{\phi^3} (V - V_c) + \frac{c_{\phi^3}}{R^2} + \ldots
\end{aligned}
\label{eq:ansatz}
\end{equation}
where $a, b, c$ are non-universal constants, and $V$ abstracts the set of pseudopotentials. Here the $c$ terms are the finite-size corrections introduced by holding the coupling fixed.
Spectra being close to conformal towers indicate $g_\phi\approx g_{\phi^3}\approx 0$, which suggests that the critical couplings drift by $1/R^2$ across system sizes.

\begin{figure*}
\includegraphics[width=0.95\textwidth]{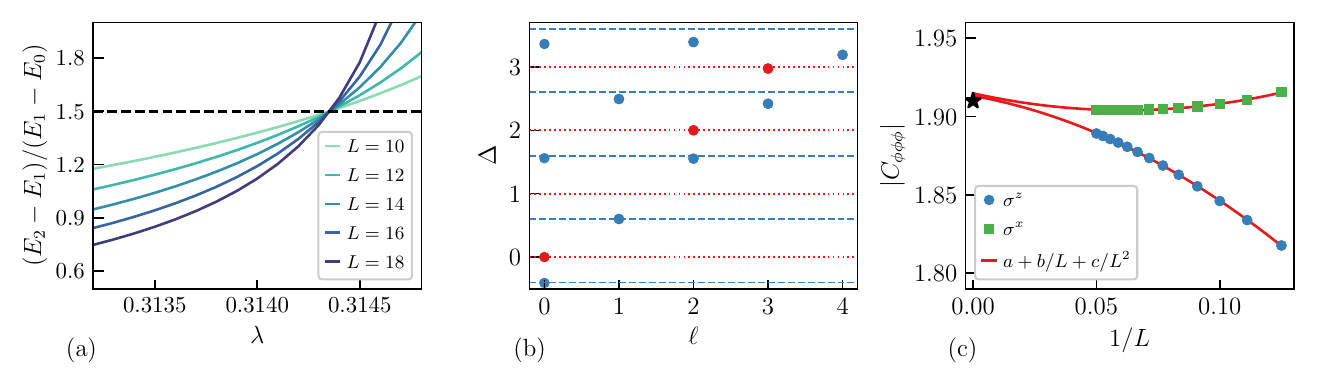}
\caption{Results of the 1+1D lattice model of Yang-Lee CFT. (a) Finite-size scaling of the ratio of energy level spacing. The black dashed line denotes the value from CFT prediction. The crossing point is $\lambda = 0.31434$ and will be used as the critical point. (b) Rescaled energy spectrum at the critical point. The system size is $L=20$. The blue and red dots correspond to the conformal family $[\phi]$ and $[\bbI]$ respectively. The lines are values from CFT prediction. States only with $0 \leq \ell \leq 4$ and $\Delta < 4$ are plotted. The spectrum at the negative momentum is obtained by a reflection thanks to the spatial inversion symmetry. (c) OPE coefficient $C_{\phi\phi\phi}$. The black star denotes the value from CFT prediction. }
\label{fig:1D YangLee}
\end{figure*}

By \eqnref{eq:ansatz}, the crossing point of finite-size scaling curves for dimensionless quantities generally corresponds to where the effective couplings align across different system sizes, rather than where they vanish. As a result, the spectrum at this crossing point often deviates from an ideal conformal tower.

In the (2+1)D Yang-Lee problem, this small drift can cause severe issues. Specifically, applying the optimized parameter at a small system size to larger ones always results in a $\calP\calT$-broken phase, which forces one to modify the optimal parameter to stay at the critical point and subsequently perform finite-size scaling analysis.
The simplest option is to fix the pseudopotentials and only modify the longitudinal field $\lambda_z$ to minimize the cost function \eqref{eq:conformality cost function}.
This can remove the relevant term $g_\phi$ and is sufficient for the scaling dimension $\Delta_\phi$ and the OPE coefficient $C_{\phi\phi\phi}$.
However, the residual irrelevant coupling $g_{\phi^3}$, owing to the non-Hermitian nature of the problem, can have a significant effect at higher energies.
For example, two nearby energies under the perturbation can collapse and gain an imaginary part.
It is therefore necessary to simultaneously modify two parameters that have overlap with $\phi$ and $\phi^3$, which we choose to be $\lambda_z$ and the pseudopotential $V_0$.

\section{Comparison with conventional methods}
\label{sec:comparison}

In this section, we demonstrate the necessity of applying the finite-size scaling method described in the previous section. We will first explain the conventional finite-size scaling method using the (1+1)D Yang-Lee CFT as an example, and then discuss the failure of the method on the fuzzy sphere.

\subsection{Conventional finite-size scaling in (1+1)D Yang-Lee problem}
\label{sec:1+1D Yang Lee}

To better appreciate the feature and subtlety of the fuzzy sphere setting, let us review the conventional method using the (1+1)D Yang-Lee problem as an instructive example.
Consider a one-dimensional Ising spin chain with the periodic boundary condition. The Hamiltonian and the $\calP\calT$ symmetry read~\cite{vonGehlen:1991zlm} 
\begin{equation}
\label{eq:1D lattice Hamiltonian}
\begin{gathered}
H = - \sum_{i} \sigma_i^x + i \lambda \sigma_i^z + V \sigma_i^z \sigma_{i+1}^z\,, \\
\calP \calT: \sigma^z \rightarrow -\sigma^z\,,\quad i \rightarrow -i\,.
\end{gathered}
\end{equation}
The phase diagram has the same structure as \figref{fig:phase diagram}, where the $\calP\calT$-symmetric and broken phases are separated by the (1+1)D Yang-Lee CFT. 
In this case, the theory has the minimal model $\calM(5,2)$ as its exact solution with only two Virasoro primaries, the conformal vacuum $\bbI$ and a $\calP\calT$-odd scalar $\phi$ with $\Delta = -2/5$~\cite{Cardy:1985yy}.
Here, $\phi$ overlaps with the scalar field in the Landau-Ginzburg Lagrangian \eqref{eq:landau ginzberg}.
The fusion rule takes the Fibonacci form $[\phi] \times [\phi] = [\bbI] + [\phi]$, and the purely imaginary OPE coefficient is $|C_{\phi\phi\phi}| = \Gamma(\frac{6}{5})^2 \Gamma(\frac{1}{5}) \Gamma(\frac{2}{5}) \Gamma(\frac{3}{5})^{-1} \Gamma(\frac{4}{5})^{-3} \approx 1.91$.

\begin{figure*}
\includegraphics[width = 0.9\textwidth]{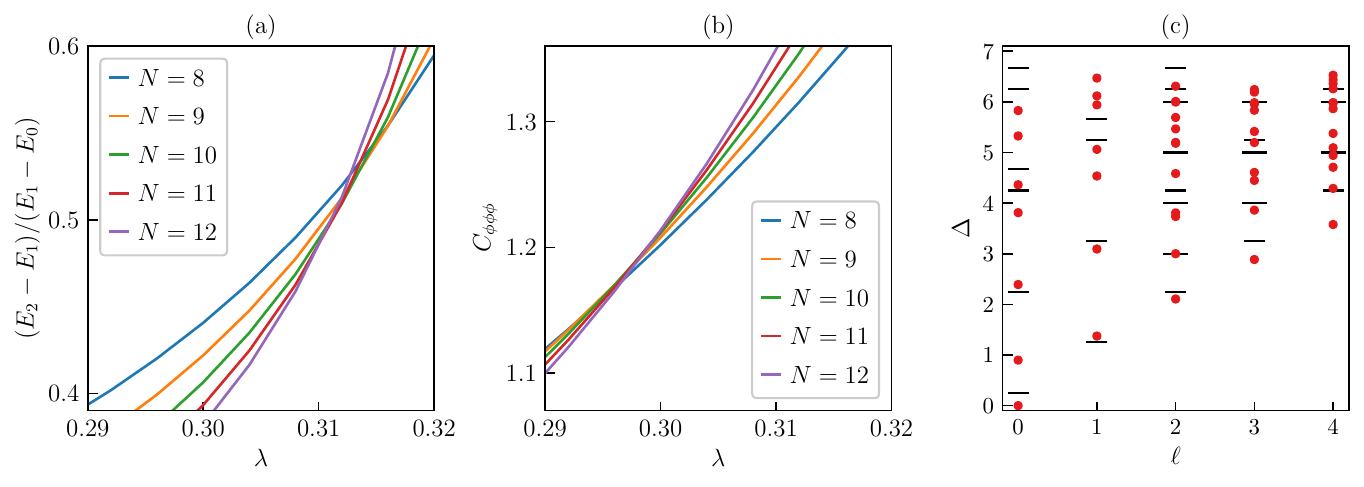}
\caption{Results of the 2+1D fuzzy sphere model of Yang-Lee CFT. (a) Finite-size scaling of the ratio of energy level spacing. The crossing is roughly at $\lambda = 0.313$. (b) Finite-size scaling of the three-point function. Here we use $n_z$. Using $n_x$ gives a similar result. (c) The rescaled spectrum at the crossing point. We use $N=12$ electrons and choose $\lambda = 0.313$. Red dots are numerical results rescaled by setting the energy of the second states in the $\ell=2$ sector to $3$. Black lines are CFT predictions for the conformal family $[\phi],[\phi^3]$ and $[T_{\mu\nu}]$ (CFT data are from the next section). For all the plots, we use the pseudopotentials $V_0 = 0.2532$, $V_1 = 0.0999$, $V_2 = -0.0461$, $V_3 = -0.0228$ that are close to ones we use to obtain the conformal tower later. }
\label{app fig:FSS fuzzysphere}
\end{figure*}

To find the critical point, we track dimensionless quantities across different system sizes as we vary the tuning parameter.
In this particular example of $\calP\calT$-symmetry breaking transition, one can also use the onset of imaginary ground state energy as a diagnosis. Here we choose to work with the crossing of dimensionless quantities as it allows for a direct comparison with the CFT prediction. 
Specifically, we choose this quantity as the ratio of energy level spacings, $(E_2 - E_1)/(E_1 - E_0)$, where $E_{0,1,2}$ denote the first three eigenenergies.
Other choices yield similar results.
One still needs to finite tune parameters to minimize the finite-size effect.
Here we find that tuning the transition along the $\lambda = V$ line leads to almost invisible finite-size effect.

The result is shown in \figref{fig:1D YangLee}(a).
As the system size increases, different curves cross at a single point, which we identify with the Yang-Lee CFT.
The ratio at this crossing precisely matches the CFT prediction. Furthermore, the energy gap at this point scales as $1/L$ with the length $L$ of the chain, confirming the gapless nature of the system. The rescaled energy spectrum, shown in \figref{fig:1D YangLee}(b), has a reasonable agreement with the expected Yang-Lee conformal tower in its low-energy part~\cite{bao2019loop}.

We can then use the ground state $\ket{\phi}$, the first excited state $\ket{\bbI}$ and the lattice operators $\sigma_i^z$ or $\sigma_i^x$ to extract the OPE coefficient $C_{\phi\phi\phi}$.
Specifically, both $\sigma_i^z$ and $\sigma_i^x$ have overlap with the two Virasoro primaries
\begin{equation}
    \sigma_i^\alpha = a_\bbI^\alpha \bbI + a_\phi^\alpha \phi + \ldots\,,
\end{equation}
where ellipses are Virasoro descendants. Neglecting descendant contributions, we estimate
\begin{equation}
|C_{\phi\phi\phi}| = \left| \frac{\braket{\phi|\sigma_i^\alpha|\phi} - \braket{\bbI|\sigma_i^\alpha|\bbI}}{\braket{\bbI|\sigma_i^\alpha|\phi}} \right|\,,
\label{eq:ope coefficient}
\end{equation}
where $\bra{\bbI}$ and $\bra{\phi}$ are the left eigenstates normalized to have a unit inner product with their right partners.
The $\calP\calT$-symmetry only constrains this OPE coefficient to be imaginary but not its sign, similar to other OPE coefficients discussed later. As a result, only the absolute value is numerically stable and meaningful in practice.
As shown in \figref{fig:1D YangLee}(c), this estimate suffers from significant finite-size corrections at small system sizes. Luckily, simply increasing the system size at the identified critical coupling can systematically reduce these effects. 
Using either $\sigma^x$ or $\sigma^z$ yields an extrapolated value that consistently matches the CFT prediction.

\subsection{Conventional finite-size scaling analysis for the (2+1)D Yang-Lee on the fuzzy sphere}
\label{app:2+1D Yang Lee FSS}

To apply the conventional finite-size scaling analysis for the fuzzy sphere Yang-Lee model, we fix the four pseudopotentials to be $V_0 = 0.2532$, $V_1 = 0.0999$, $V_2 = -0.0461$, $V_3 = -0.0228$, which is essentially the one obtained through the gradient descent algorithm, and we use the imaginary longitudinal field as the tuning parameter. 
We examine how dimensionless quantities change with the tuning parameter across different system sizes. 
There are multiple choices. Here we show results for the ratio of the energy level spacings and the three-point function for a better comparison with the results obtained in the (1+1)D Yang-Lee problem and through the other method in (2+1)D.

We choose the first three eigenenergies $E_0, E_1, E_2$ that correspond to the conformal vacuum, $\phi$ and $\partial\phi$ in the true Yang-Lee CFT. 
The result of the ratio $(E_2 - E_1)/(E_1 - E_0)$ is shown in \figref{app fig:FSS fuzzysphere}~(a).
One can also compute the three-point function, or more precisely the quantity \eqnref{eq:ope coefficient}, and the result is shown in \figref{app fig:FSS fuzzysphere}~(b).
Curves at different system sizes do cross each other but the crossing point keeps drifting as the system size increases. In practice, we are not able to find parameters that remove this drifting. 

We can use the logic explained in \secref{sec:FSS} to understand this observation more quantitatively. For simplicity, we do the analysis for the ratio of level spacings. 
Let us assume that the system is not far from the true critical point so that we can use the effective description \eqnref{eq:perturbation Hamiltonian}. 
We can then apply the conformal perturbation to estimate the ratio of level spacing by
\begin{equation*}
    \frac{E_2 - E_1}{E_1 - E_0} \approx \frac{1 + \tilde{g}_\phi R^{3 - \Delta_\phi} A(\Delta_\phi) + \tilde{g}_{\phi^3} R^{3 - \Delta_{\phi^3}} A(\Delta_{\phi^3})}{\Delta_\phi + \tilde{g}_\phi R^{3 - \Delta_\phi} + \tilde{g}_{\phi^3} R^{3 - \Delta_{\phi^3}}}
\end{equation*}
where $A(\Delta) = \Delta(\Delta - 3)/6\Delta_\phi$ is obtained via conformal algebra, $\tilde{g}_\phi = 4\pi |C_{\phi\phi\phi}| g_\phi$ and $\tilde{g}_{\phi^3} = 4\pi |C_{\phi\phi\phi^3}| g_{\phi^3}$.
Due to the irrelevance of $\phi^3$ and $|C_{\phi\phi\phi^3}|$ being small, we can ignore the contribution from $\phi^3$ and approximately have
\begin{equation*}
     \frac{E_2 - E_1}{E_1 - E_0} \approx \frac{1}{\Delta_\phi} \left(1 + \frac{\Delta_\phi^2 - 3\Delta_\phi - 6}{6} \tilde{g}_\phi R^{3 - \Delta_\phi} \right) 
\end{equation*}
Using the ansatz formula \eqref{eq:ansatz} for the relation between $g_\phi$ and the microscopic Hamiltonian
\begin{equation*}
\frac{E_2 - E_1}{E_1 - E_0} \approx \frac{1}{\Delta_\phi} \left(1 - a ( \lambda - \lambda_c) R^{3 - \Delta_\phi} + c R^{1 - \Delta_\phi} \right)\,,
\end{equation*}
where we further drop the pseudopotential term by assuming it is already at the critical value in the thermodynamic limit (we justify it in the next section). In our convention $a>0$ and $c$ has an undetermined sign. 
The presence of the ambiguity term is responsible for the observed drift of the crossing point and its effect is pronounced especially when $\Delta_\phi < 1$, which is the regime the non-unitary Yang-Lee CFT sits at.

If we compare the behavior of the dimensionless quantity with what we have shown for the (1+1)D lattice model, the result here has a much worse crossing behavior.
Furthermore, the values of the dimensionless quantities at the crossing points are much smaller than those obtained via the other method shown in the next section. Indeed, the rescaled energy spectrum at this approximate crossing point, shown in \figref{app fig:FSS fuzzysphere}~(c), is far from being a conformal tower.
We believe this discrepancy to be intrinsic to the fuzzy sphere setup, and the method presented in \secref{sec:FSS} is necessary.

\begin{figure*}[!t]
	\centering
	\includegraphics[width=4.3cm]{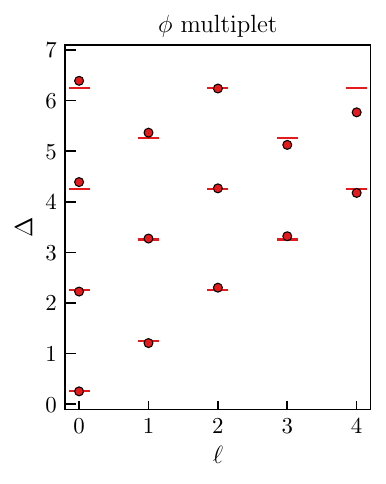}
	\includegraphics[width=4.3cm]{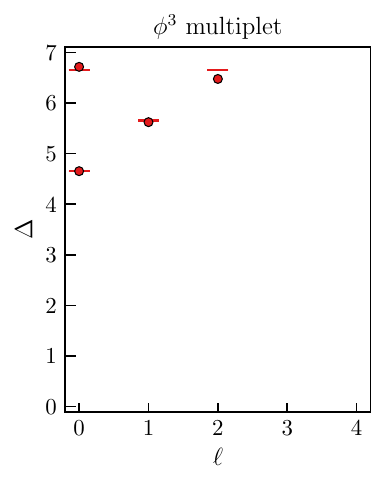} \\
	\includegraphics[width=4.3cm]{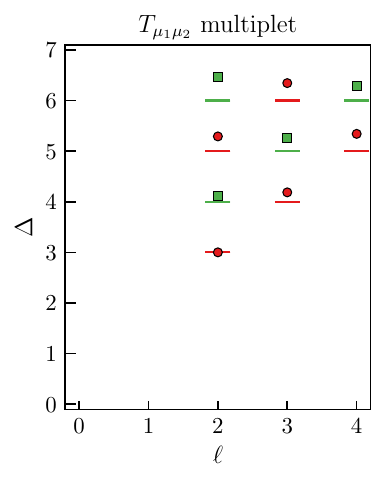}
	\includegraphics[width=4.3cm]{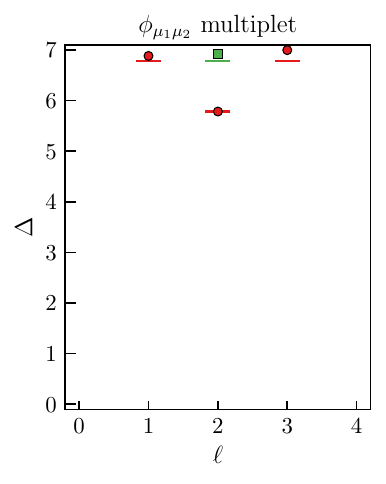}
	\includegraphics[width=4.3cm]{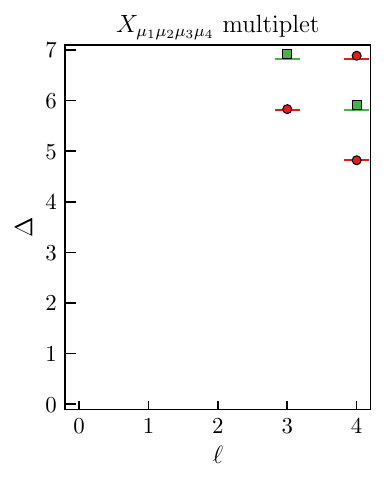}
	\caption{Conformal multiplet of low-lying primary operators. In each plot the lowest energy state is the primary and the rest are descendants. Red dots and green squares are numerical results for states of even and odd parity, respectively. Lines are integer shifts of the energy of the primary state. Due to the finite-size effect, the energy of the level-2 spin-4 descendant of $X_{\mu_1\mu_2\mu_3\mu_4}$ has a small imaginary part ($0.05i$). Here we only plot its real part. States of $\ell\geq 4$ or $\Delta>7$ are subjected to large finite-size effects and are not shown here.}
	\label{fig:conformal multiplet}
\end{figure*}

\section{Results}
\label{sec:result}

Now, we present the results obtained by using the method described in \secref{sec:FSS}. 
The output parameter from the gradient descent can produce good conformal spectra even at small system sizes, e.g., with $N=8$ electrons.
Using larger system sizes does give a better result.
Here, we fix $V_1$, $V_2$ and $V_3$ obtained from the gradient descent in a small system. Then, we modify $\lambda_z$ and $V_0$ by minimizing the cost function \eqref{eq:conformality cost function} as we increase the system size.
Gradient descent with $N=12$ electrons yields optimal parameters, $V_0 = 0.244196$, $V_1 = 0.0999$, $V_2 = -0.0461$, $V_3 = -0.0228$ and $\lambda_z = 0.33603$, which together produce the conformal spectrum shown in \figref{fig:conformal multiplet}.
These optimal parameters are not unique. There are multiple combinations that pass the conformal perturbation check, and all valid sets yield consistent conformal data.

\figref{fig:conformal multiplet} shows five low-lying primary states including the stress-energy tensor.
The first excited state corresponds to the only relevant scalar primary $\phi$ and has an energy $\Delta = 0.25$, which clearly violates the unitarity bound.
That $\phi$ has such a small scaling dimension implies that the outcome is sensitive to the precise parameter values, and retaining high numerical precision is crucial to reach precise results.
Additionally, we identify the irrelevant scalar $\phi^3$ at the energy $\Delta = 4.67$ and the spin-4 primary $X_{\mu_1\mu_2\mu_3\mu_4}$ at $\Delta = 4.82$, which are known from bootstrap studies. 
Notably, we also identify a spin-2 primary $\phi_{\mu_1\mu_2}$ at $\Delta = 5.78$ that has not been found previously.
Beyond the plotted regime, we find two complex conjugate eigenenergies at $\Delta \approx 7.8$ in the spinless sector. One possible explanation of this pair of states is the collapsing of the level-2 descendant $\Box\phi^3$ and an unknown primary scalar around $8$, which then approximately sets a lower bound on the scaling dimension of the next irrelevant scalar primary in the spectrum. We will use this number in our finite-size scaling analysis later.

\begin{figure}[!t]
	\includegraphics[width=0.47\textwidth]{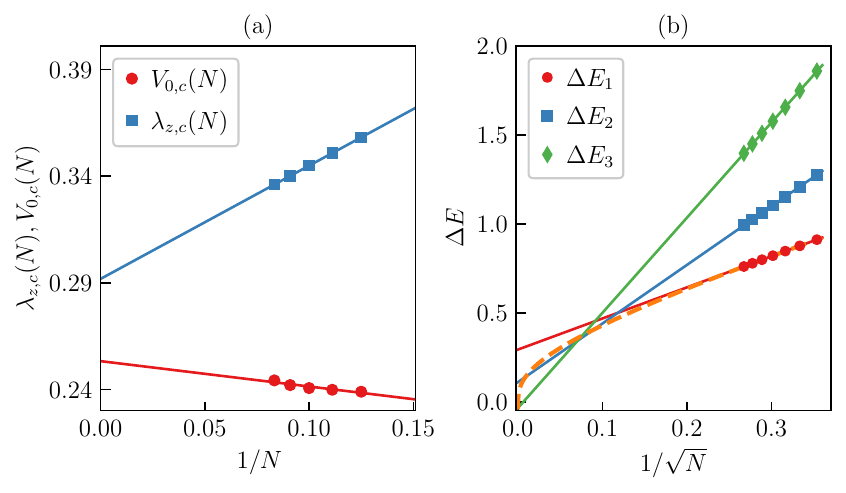}
	\caption{(a) Drift of the critical coupling as a function of the system size $N=8,9,\ldots,12$. The extrapolated critical couplings in the thermodynamic limit are $\lambda_c = 0.2919$ and $V_{0,c} = 0.2532$. (b) Vanishing of the energy gap in the thermodynamic limit. Here $\Delta E_{i=1,2,3}$ are the energy gaps of the first, second and third excited states. The system sizes are $N=8,9,\ldots,14$. For both plots we choose $V_1 = 0.0999$, $V_2 = -0.0461$ and $V_3 = -0.0228$.}
	\label{fig:finite-size scaling}
\end{figure}

The sets of $\lambda_{z,c}(N)$ and $V_{0,c}(N)$ that yield the most conformal spectrum with fixed $V_1$, $V_2$, $V_3$ at different system sizes are shown in \figref{fig:finite-size scaling}~(a). 
A clear systematic drift in the optimal parameters is observed. 
One can fit it well to a $1/N$ dependence (equivalent to $1/R^2$), which is what we expect from \eqnref{eq:ansatz}.
As a further consistency check, we simulate the system at the extrapolated critical couplings $\lambda_{z,c}$ and $V_{0,c}$ in the thermodynamic limit and examine the energy gap as a function of the system size, shown in \figref{fig:finite-size scaling}~(b).
Using the simplest linear fit in $1/\sqrt{N}$, the energy gap of the first excited state asymptotes to a small but nonzero value. 
According to the analysis in the previous section, we can attribute this deviation to the presence of $\phi$ when using $\lambda_{z,c}$ and $V_{0,c}$ in a finite-size system. 
Recalling \eqnref{eq:ansatz}, the extra $c_\phi$ term shifts the energy at the leading order by
\begin{equation}
    \delta E = \frac{c_\phi}{R^2} \langle \phi | \int \phi(x) d^2 x |\phi \rangle \propto R^{-\Delta_\phi}\,,
\end{equation}
which overwhelms the $1/R$ piece at large $R$ for $\Delta_\phi < 1$.
Accounting for this perturbation, the expected scaling form becomes $\Delta E_1(N) = a_0 + a_1/\sqrt{N} + a_2/N^{\Delta_\phi/2}$. This improved fit, shown as the orange dashed curve, correctly extrapolates to zero in the thermodynamic limit.
On the other hand, the energy gaps of the next two excited states asymptote to values much closer to zero under the linear fitting.
This is possibly due to the quantitative fact that the $\phi$ perturbation tends to have a smaller effect for the descendant states.

\begin{figure}[!t]
	\includegraphics[width=0.47\textwidth]{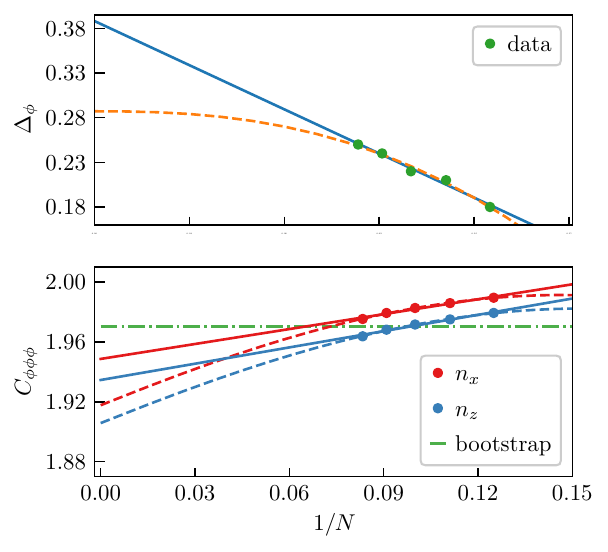}
	\caption{(Upper) Finite-size scaling of the scaling dimension $\Delta_\phi$. (Lower) Finite-size scaling of the OPE coefficient $C_{\phi\phi\phi}$. 
    In the two plots, the dots are the numerical data, the solid lines are linear fitting in $1/N$ and the dashed lines are non-linear fitting, $a + b N^{-5/2}$ for the scaling dimension and $a + b N^{-1} + cN^{-5/2}$ for the OPE coefficients. The linear fit is added just for a comparison.
    The $1/N$ term comes from the leading scalar descendant, the non-linear term comes from an irrelevant scalar primary, the power of which is determined by the estimated low bound of its scaling dimension $\Delta\approx 8$ as $(3-\Delta)/2 \approx 5/2$.}
	\label{fig: phi FSS}
\end{figure}

We then use the system-size dependent optimal parameters to perform a finite-size scaling analysis of key CFT data, the scaling dimensions and the OPE coefficients.
\figref{fig: phi FSS} illustrates the result for two of the simplest quantities: the scaling dimension $\Delta_\phi$ and the absolute value of the OPE coefficient $C_{\phi\phi\phi}$.
Similar to the critical couplings, $\Delta_\phi$ also exhibits a systematic drift with the system size. 
As the two-parameter fine-tuning already removes the effect of $\phi$ and $\phi^3$, we expect the drift in $\Delta_\phi$ to come from other residual irrelevant perturbations. 
Based on the behavior of the spectrum, we have estimated that the next irrelevant scalar primary should have a scaling dimension around at least 8. This estimate is then used in a nonlinear finite-size fitting, as shown by the dashed curve in the figure. 
The final extracted result is $\Delta_\phi = 0.29$, which is slightly larger than what has been reported via other methods. Using a larger lower bound on the scaling dimension of the next irrelevant primary for the non-linear fitting will yield smaller results that are closer to the previous findings.

To extract OPE coefficients, we consider the microscopic operators $n_z(\Omega)$ and $n_x(\Omega)$. Similar to the (1+1)D example, both of them have nonzero overlaps with all primary operators
\begin{equation}
	\label{eq:operator overlap}
	n_\alpha(\Omega) = a_\alpha^\bbI \bbI + a_\alpha^\phi \phi + a_\alpha^{\phi^3} \phi^3 + \ldots
\end{equation}
where the coefficient $a_\alpha^{O}$ can be estimated via calculating $\braket{\bbI|n_\alpha(\Omega)|O}$ for a scalar primary $O$, and ellipses represents descendants and higher primaries.
We found that $a_\alpha^{\bbI}$ and $a_\alpha^{\phi}$ are larger than the coefficients of other primaries by at least one order of magnitude.
Therefore it is a reasonable approximation to truncate \eqnref{eq:operator overlap} to the first two terms and calculate the OPE coefficients~\cite{Hu:2023xak}. 
Specifically, we compute
\begin{equation}
    |C_{\phi O_1 O_2}| =  \Big| \frac{\langle O_1 |n_\alpha| O_2 \rangle - a_\alpha^\bbI\langle O_1 |O_2 \rangle}{\langle \bbI |n_\alpha| \phi \rangle} \Big|\,.
    \label{eq:CphiO1O2}
\end{equation}
Up to subleading correction, \eqnref{eq:CphiO1O2} agrees with the ordinary definition of the OPE coefficient for scalar $O_1,O_2$ and differs by a proportionality constant if they are spinful.
The result of $C_{\phi\phi\phi}$ is shown in \figref{fig: phi FSS} as an example. The systematic drift has two contributions: (1) $n_\alpha$ having overlap with the scalar descendant, (2) the eigenstates receiving perturbation from residual irrelevant perturbation. We can include both effects via a non-linear fitting and the result is shown by the dashed lines. The extrapolated result is between $1.91$ and $1.92$, which is slightly smaller than the bootstrap result.

We summarize our results obtained through the finite-size scaling analysis in \tabref{tab:scaling dimension} and \ref{tab:OPE coefficients}, where we do not exhaust all the accessible OPE coefficients.
Notably we found $|C_{\phi\phi\phi^3}| \ll |C_{\phi\phi\phi}|$, which justifies the approximation of ignoring $[\phi^3]$ contribution adopted by the bootstrap approach in Ref.~\cite{Gliozzi:2013ysa}.

\begin{table}[t!]
	\centering
	\begin{tabularx}{0.45\textwidth}{C|C|C|C|C}
		\hline
		\hline
		Operator & $\phi$ & $\phi^3$ & $\phi_{\mu_1\mu_2}$ & $X_{\mu_1\mu_2\mu_3\mu_4}$ \\
		\hline
		FS & 0.29 & 4.71 & 5.83 & 4.90 \\
		\hline
		CB & 0.235 & 5.0 & NA & 4.75 \\
		\hline
		\hline
	\end{tabularx}
	\caption{Scaling dimensions of the low-lying primary operators. The scaling dimension of $\phi$ has also been estimated by other methods. For the reader's convenience we list the results collected by Ref.~\cite{Gliozzi:2013ysa} here: 0.214 (Ising with imaginary field), 0.242 (Fluids), 0.222 (Animal), 0.220 - 0.25 ($\varepsilon$-expansion).}
	\label{tab:scaling dimension}
\end{table}
\begin{table}[t!]
	\centering
	\begin{tabular}{p{3cm} p{2cm} p{1cm}}
		\hline
		\hline
		OPE coefficients & FS & CB\\
		\hline
		$|C_{\phi\phi\phi}|$ & $1.91-1.92$ & 1.97 \\
        $|C_{\phi\phi\phi^3}|$ & $0.02-0.16$ & NA \\
		$|C_{\phi\phi^3\phi^3}|$ & $2.66-2.67$ & NA \\
		$|C_{\phi\phi T}|$ & $0.03 - 0.04$ & 0.08 \\
        $|C_{\phi T T}|$ & $2.04 - 2.12$ & NA \\
		$|C_{\phi\phi\phi_{\mu_1\mu_2}}|$ & $0.005 - 0.08$ & NA \\
        $|C_{\phi\phi_{\mu_1\mu_2}\phi_{\mu_1\mu_2}}|$ & $1.56 - 1.64$ & NA \\
        $|C_{\phi\phi X}|$ & $0 - 0.03$ & NA \\
        $|C_{\phi X X}|$ & $1.465 - 1.473$ & NA \\
		\hline
		\hline
	\end{tabular}
	\caption{Operator product expansion coefficients of the low-lying primaries. The result obtained from using $n_x$ and $n_z$ generally do not agree with each other and provides us with an error estimation. Here, $T$ and $X$ stands for $T_{\mu_1\mu_2}$ and $X_{\mu_1\mu_2\mu_3\mu_4}$.}
	\label{tab:OPE coefficients}
\end{table}

\section{Discussion}
\label{sec:discussions}

In this work we apply the fuzzy sphere regularization to study the (2+1)D Yang-Lee CFT and identify five non-identity primary fields.
The extracted conformal data are largely consistent with results obtained by other methods where available.
However, there is one subtlety of the spin-4 primary $X_{\mu_1\mu_2\mu_3\mu_4}$. In practice, there is no sharp way to distinguish this state from another one at a higher energy around $\Delta = 5.69$ since using either of them yields a nice conformal tower. We take the bootstrap result as guidance and assume that the one with the lower energy is the primary.
It will be useful to construct the special conformal generator to further distinguish them.
However, naively applying the prescription in \cite{Fardelli:2024qla, Fan:2024vcz} fails, and we leave a more careful treatment to further work.

It is natural to further examine the associated defect CFT and the (violation of) $F$-theorem~\cite{Zhou:2023fqu, Hu:2023ghk,Cuomo:2024psk,Hu:2024pen}.
It is also interesting to further explore other non-unitary CFTs, especially the putative complex CFT conjectured to be near the deconfined quantum critical point~\cite{Walking1,Walking2,Zhou:2023qfi}. 
Another interesting task is to use the four-dimensional quantum Hall system as a potential regularization to study the (4+1)-dimensional Yang-Lee CFT~\cite{Zhang:2001xs}.

We also systematically discuss a finite-size scaling analysis tailored to the fuzzy sphere.
This method is essential for accurately analyzing our Yang-Lee model and may prove broadly useful for other CFTs on the fuzzy sphere.
For comparison, we revisit conventional finite-size scaling in the (1+1)D Yang-Lee model in which, as a bonus, we identify a special parameter where finite-size effects are remarkably suppressed.

One limitation of the current finite-size scaling method is its reliance on solving for many excited states, which becomes computationally demanding for large systems.
It would be both conceptually interesting and numerically more efficient to explore alternative approaches that identify the critical point using only the ground state wavefunction, enabling analysis at larger system sizes.

\bigskip

\emph{Note added:} Near the completion of this work, we become aware of \cite{Grisha:toappear, Joan:toappear}, which studies the (2+1)D Yang-Lee CFT on the fuzzy sphere using a different approach.

\section*{Acknowledgment}

We would like to thank Zhehao Dai, Daniel Parker, Zlatko Papi\'c, Slava Rychkov, Cristian Voinea, Jie Wang, Taige Wang, Xueda Wen, Yantao Wu and Michael Zaletel for insightful discussions. 
We thank Yantao Wu for kindly allowing us to use his workstation.
R.F. thanks Zlatko Papi\'c and Cristian Voinea for collaborations on related projects and helpful discussion on the finite-size scaling.
We are especially grateful to Yin-Chen He for providing helpful suggestions throughout the entire project.
R.F. is supported by the Gordon and Betty Moore Foundation (Grant GBMF8688). This research was supported in part by grant NSF PHY-2309135 to the Kavli Institute for Theoretical Physics (KITP).
A.V. and J.D. were supported by the grant NSF DMR-2220703.

\onecolumngrid
\appendix
\setcounter{secnumdepth}{2}

\twocolumngrid
\bibliography{ref.bib}

\end{document}